\documentclass[a4paper,11pt,final,twocolumn]{article}

\usepackage{balance}		
\usepackage{graphicx}		
\usepackage{times}			
\usepackage{url}			
\usepackage{dblfloatfix}	
\usepackage{amsmath} 
\usepackage{amssymb} 
\usepackage{mathrsfs}
\usepackage{authblk}
\usepackage[numbers]{natbib}

\makeatletter
\def\url@leostyle{%
  \@ifundefined{selectfont}{\def\UrlFont{\sf}}{\def\UrlFont{\small\bf\ttfamily}}}
\makeatother
\def\pprw{8.5in}
\def\pprh{11in}

\usepackage[pdftex]{hyperref}
\hypersetup{
	bookmarksnumbered,
	pdfstartview={FitH},
	colorlinks,
	citecolor=blue,
	filecolor=blue,
	linkcolor=blue,
	urlcolor=blue,
	breaklinks=true
}

\begin{document}

\title{Straightness of rectilinear vs. radio-concentric networks: modeling, simulation and comparison}

\author[1,2]{Didier Josselin\thanks{\href{mailto:didier.josselin@univ-avignon.fr}{didier.josselin@univ-avignon.fr}}}
\author[2]{Vincent Labatut\thanks{\href{mailto:vincent.labatut@univ-avignon.fr}{vincent.labatut@univ-avignon.fr}}}
\author[3]{Dieter Mitsche\thanks{\href{mailto:dmitsche@unice.fr}{dmitsche@unice.fr}}}
\affil[1]{UMR ESPACE 7300, CNRS, Universit\'e d'Avignon, France}
\affil[2]{Laboratoire Informatique d'Avignon EA 4128, Universit\'e d'Avignon, France}
\affil[3]{Laboratoire de Math\'ematiques Dieudonn\'e, Universit\'e de Nice, France}

\renewcommand\Authands{ and }

\date{}

\maketitle

\begin{abstract}
This paper proposes a comparison between rectilinear and radio-concentric networks. Indeed, those networks are often observed in urban areas, in several cities all over the world. One of the interesting properties of such networks is described by the \textit{straightness} measure from graph theory, which assesses how much moving from one node to another along the network links departs from the network-independent straightforward path. We study this property in both rectilinear and radio-concentric networks, first by analyzing mathematically routes from the center to peripheral locations in a theoretical framework with perfect topology, then using simulations for multiple origin-destination paths. We show that in most of the cases, radio-concentric networks have a better straightness than rectilinear ones. How may this property be used in the future for urban networks?
\end{abstract}

\section{Introduction}
We propose to study and to compare the \textit{straightness} of two very common networks: rectilinear and radio-concentric networks. This measure, also called directness, is related to another index called efficiency \cite{Vragovic2005}, and is the reciprocal of the tortuosity (a.k.a. circuity) \cite{Kansal2001} -- see section \ref{sec:StrDef} for a formal definition. In a general meaning, a high straightness reflects the capacity of a network to enable the shortest routes. It is somehow an accessibility assessment: the higher the straightness, the shorter the distance (or moving time) on the network, due to reduced detours.

First, we show a few pictures of old and current networks observed in the real-world, to highlight their peculiar structures \cite{man04}. Hippodamian or rectilinear (also called Manhattan) networks are made of rectangular polygons, while radio-concentric networks show a center location and a series of radial and circular links. 

Second, we model theoretical rectilinear and radio-concentric networks. We introduce the mathematical formula of the straightness for center-to-periphery routes, and we demonstrate that in most of the cases, even with a low number of rays, radio-concentric networks provide straighter paths. 

Third, we empirically process the straightness of all routes (any node to any node) using Dijkstra's shortest path algorithm \cite{dijk71}. The results confirm that for any type of routes, radio-concentric structure has higher straightness.

\subsection{Some Rectilinear Networks}
\begin{figure}[!h]
	\centering
	\includegraphics[width=0.8\columnwidth]{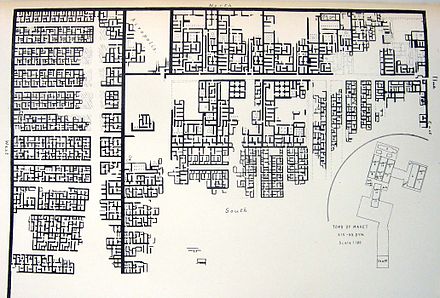}
	\caption{The antique Egyptian city of Kahun (pyramid of Sesostris); Wikipedia.}
	\label{fig:kahun}
\end{figure}

Rectilinear networks are very common all over the world. They are also called Hippodamian maps due to the Greek architect Hippodamos. These networks are characterized by road sections crossing at right angles. Built according to land registry and road construction efficiency in a process of urban sprawl, they look like pure theoretical shapes: a grid of squares of identical surface and side length. Figure~\ref{fig:kahun} shows an example of an old hippodamian urban structure in the antique Egyptian city of Kahun (pyramid of Sesostris).

\begin{figure}[!h]
	\centering
	\includegraphics[width=0.8\columnwidth]{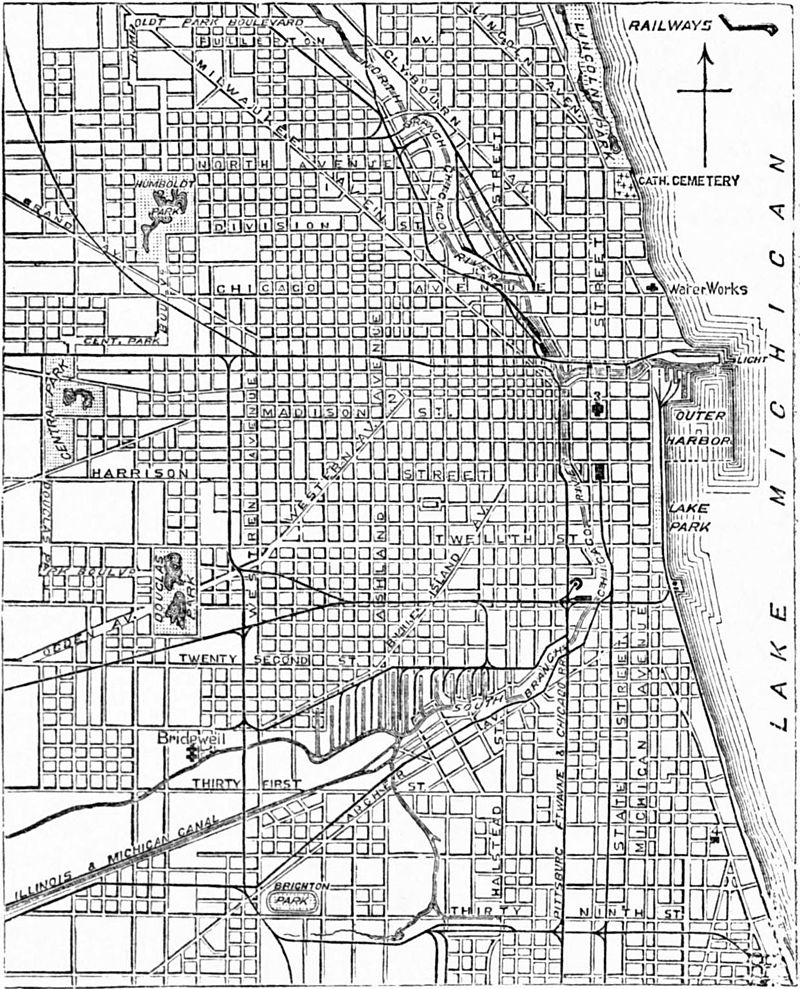}
	\caption{Chicago (USA) in 1848; Wikipedia.}
	\label{fig:chicago}
\end{figure}

\begin{figure}[!h]
	\centering
	\includegraphics[width=0.8\columnwidth]{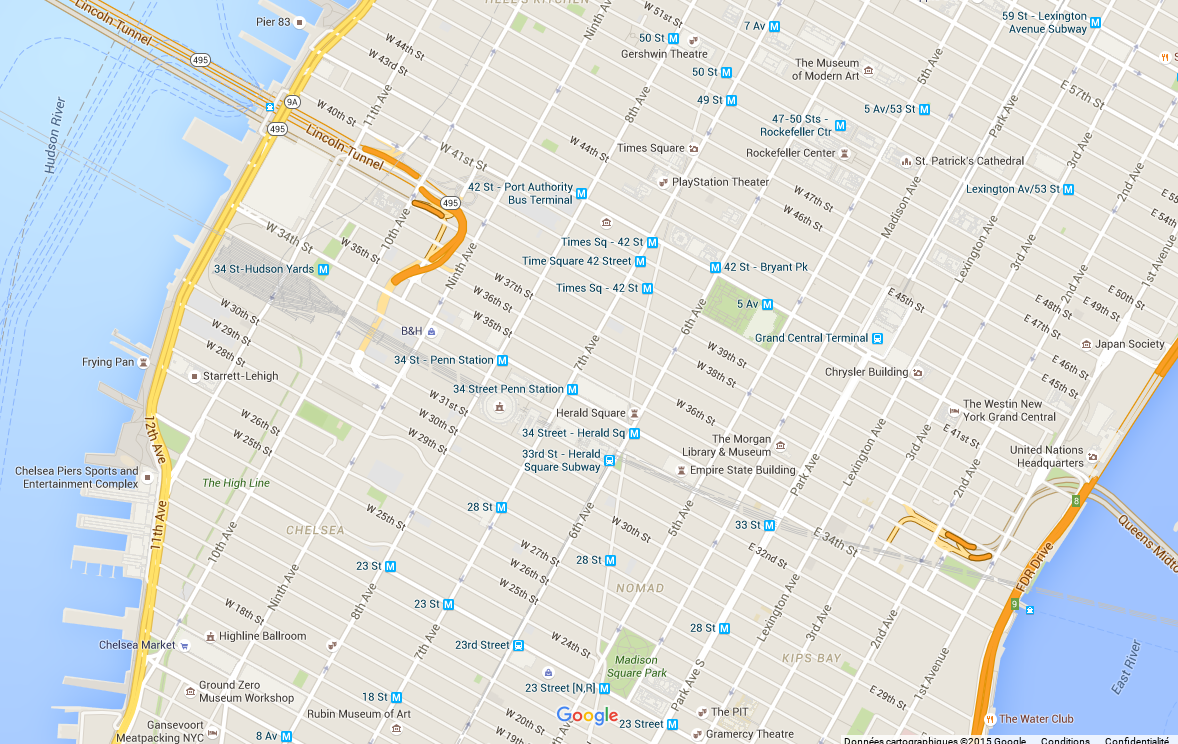}
	\caption{The famous rectilinear network of Manhattan (USA); Google Maps.}
	\label{fig:manhattan}
\end{figure}

\begin{figure}[!h]
	\centering
	\includegraphics[width=0.8\columnwidth]{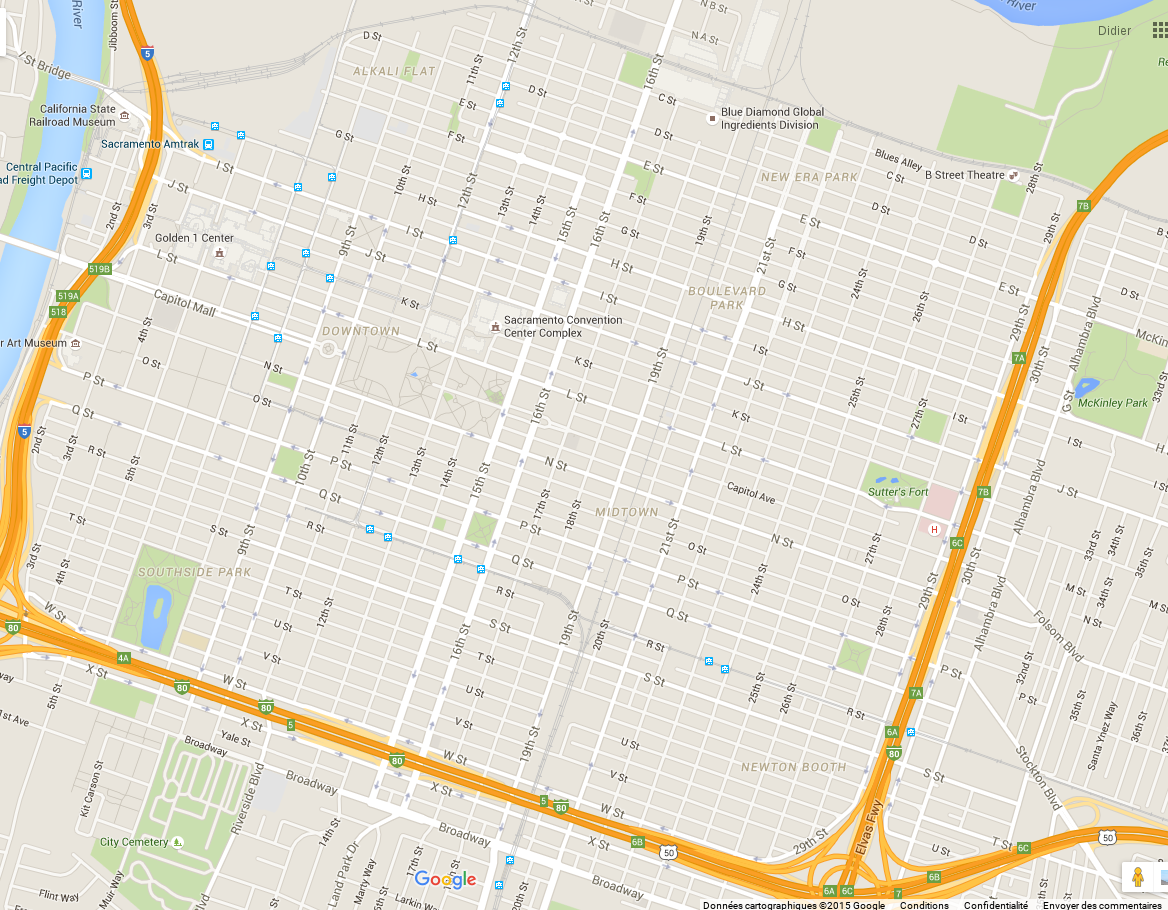}
	\caption{The very regular road network of Sacramento (USA); Google Maps.}
	\label{fig:sacramento}
\end{figure}

\begin{figure}[!h]
	\centering
	\includegraphics[width=0.8\columnwidth]{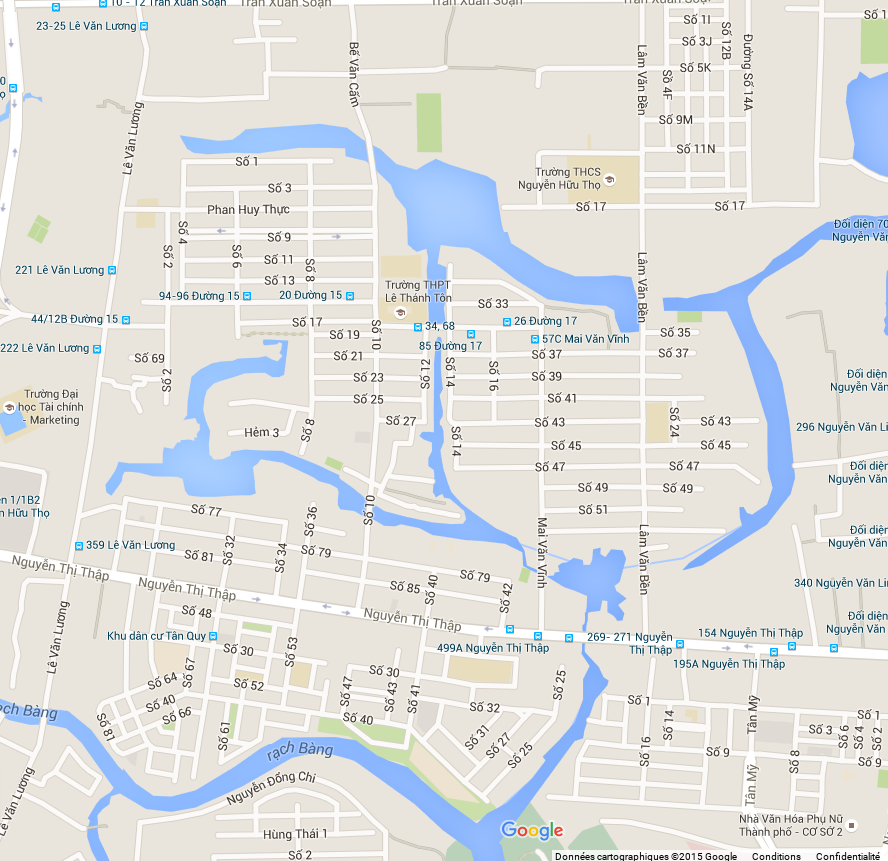}
	\caption{Road network of H\^o Chi Minh (Vietnam); Google Maps.}
	\label{fig:hochiminh}
\end{figure}

More recent rectilinear networks are illustrated by the American cities of Chicago in 1848 (Figure~\ref{fig:chicago}), New-York with Manhattan (Figure~\ref{fig:manhattan}), Sacramento (Figure~\ref{fig:sacramento}) and also the Asian city of H\^o Chi Minh in Vietnam (Figure~\ref{fig:hochiminh}). 

\subsection{Radio-concentric Networks and Shapes}
The second type of network we study is also very common. It is very different in its shape, although it presents basic polygonal entities of various sizes, due to the radial structure. In these networks, the center is somehow a fuzzy location, that often refers to the old part of the town. From this supposed center, radial roads are drawn, crossing a series of perpendicular ring roads, depending on the surface of the city.

\begin{figure}[!h]
	\centering
	\includegraphics[width=0.8\columnwidth]{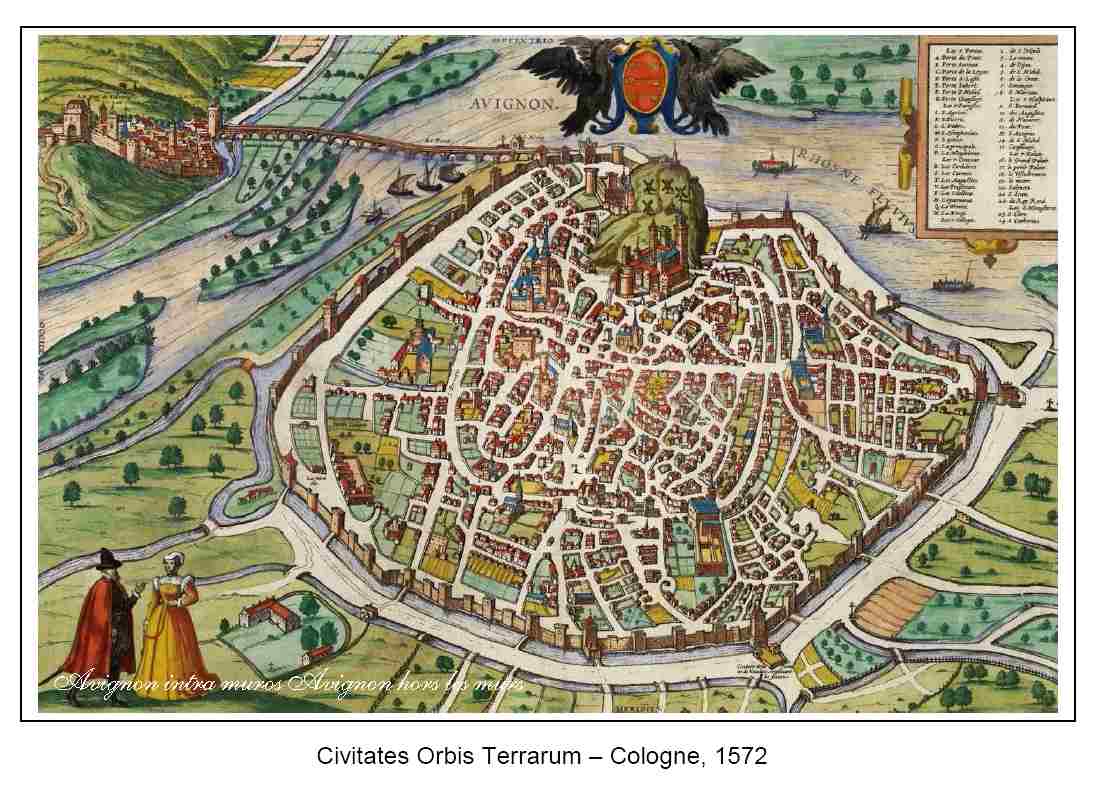}
	\caption{Medieval map of \textit{intramuros} Avignon (France).}
	\label{fig:avignon}
\end{figure}

\begin{figure}[!h]
	\centering
	\includegraphics[width=0.8\columnwidth]{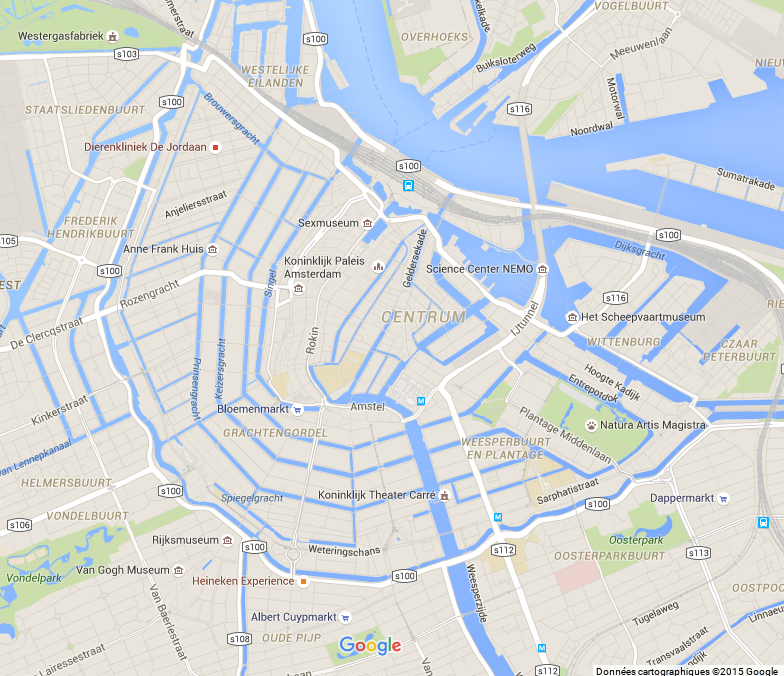}
	\caption{Amsterdam (The Netherlands); Google Maps.}
	\label{fig:amsterdam}
\end{figure}

\begin{figure}[!h]
	\centering
	\includegraphics[width=0.8\columnwidth]{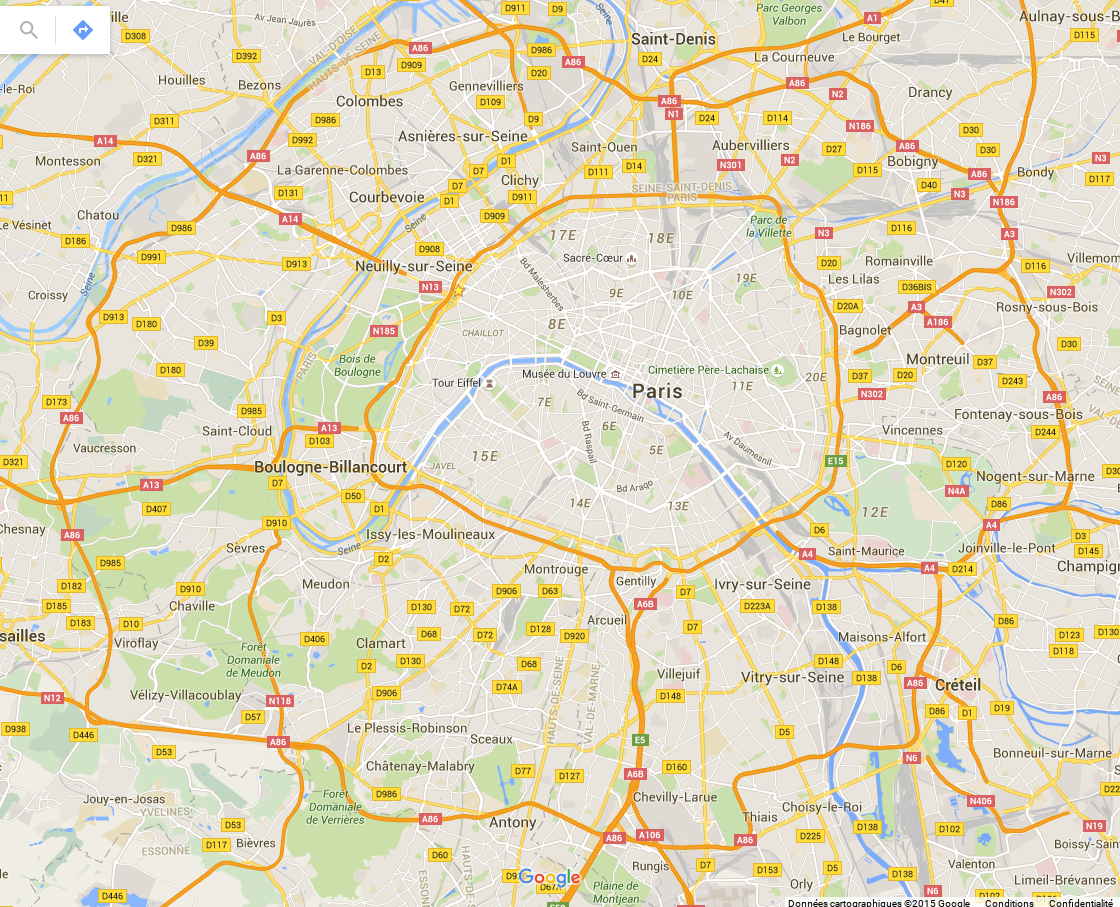}
	\caption{Paris (France); Google Maps.}
	\label{fig:paris}
\end{figure}

\begin{figure}[!h]
	\centering
	\includegraphics[width=0.8\columnwidth]{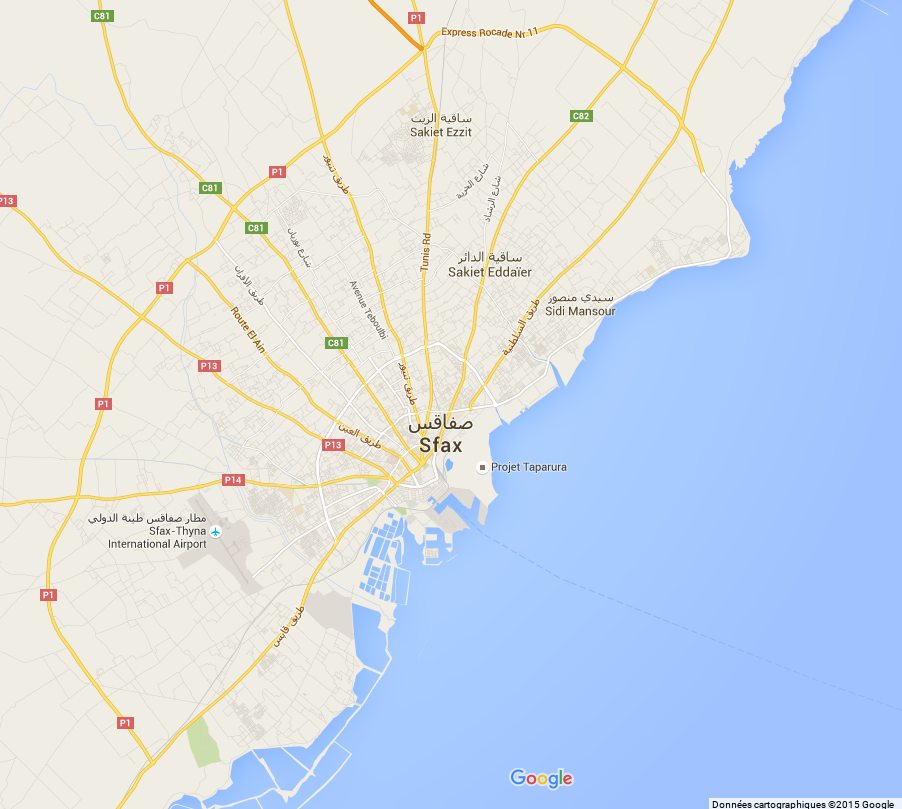}
	\caption{Sfax (Tunisia); Google Maps.}
	\label{fig:sfax}
\end{figure}

\begin{figure}[!h]
	\centering
	\includegraphics[width=0.8\columnwidth]{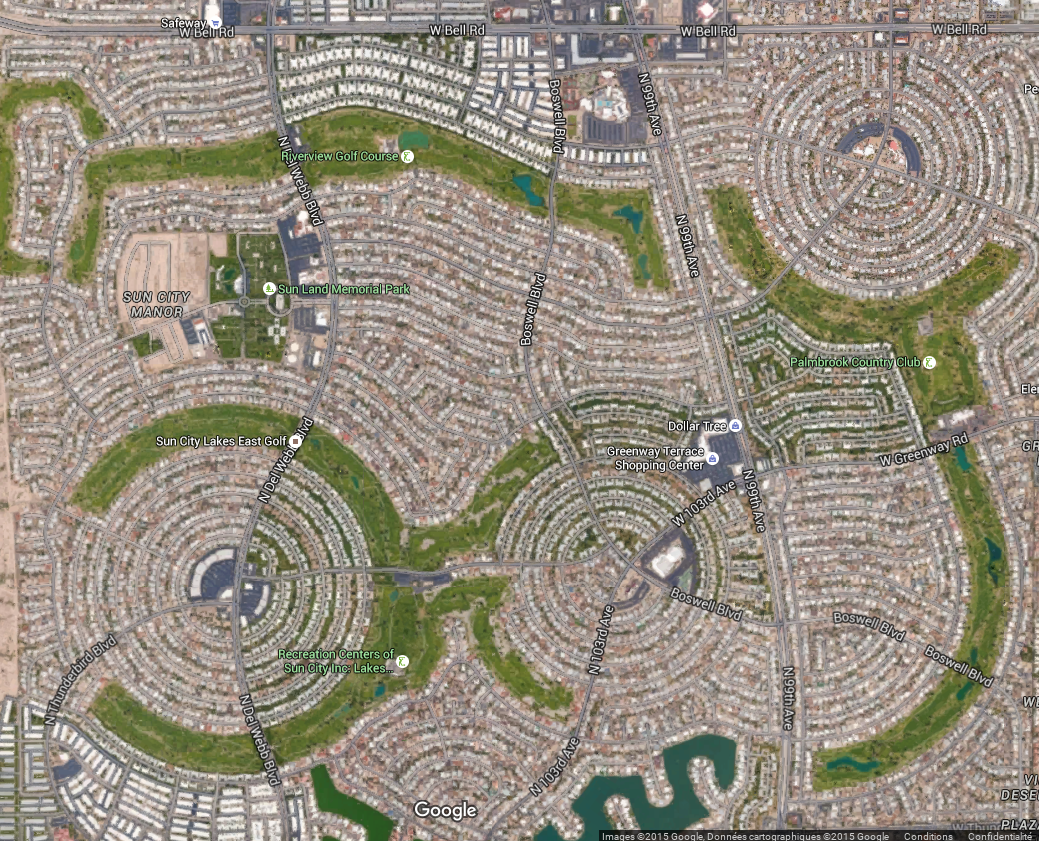}
	\caption{Suncity in Arizona (USA); Google Maps.}
	\label{fig:suncity}
\end{figure}

\begin{figure}[!h]
	\centering
	\includegraphics[width=0.8\columnwidth]{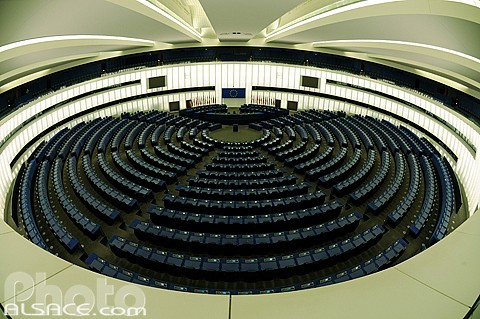}
	\caption{European Parliament in Strasbourg (France); Photo-alsace.com.}
	\label{fig:europe}
\end{figure}

\begin{figure}[!h]
	\centering
	\includegraphics[width=0.8\columnwidth]{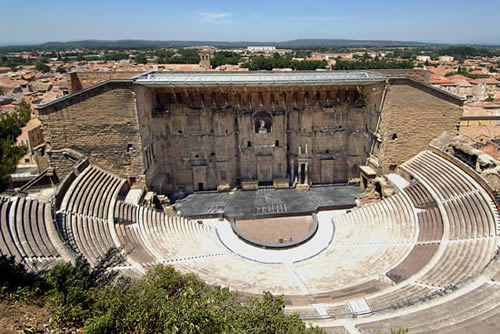}
	\caption{Ancient theater of Orange (France); Avignon-et-provence.com.}
	\label{fig:orange}
\end{figure}

These shapes were already visible in old maps such as medieval Avignon (cf. Figure~\ref{fig:avignon}), especially in its $intramuros$ part, which is surrounded by battlements. In more recent urbanization, many urban areas reveal concentric shapes. Figures~\ref{fig:amsterdam}, \ref{fig:paris}, \ref{fig:sfax}, \ref{fig:suncity}, \ref{fig:europe} and \ref{fig:orange} are good examples of how much geometrical these shapes are. Indeed, as with rectilinear networks, there exist many degrees of spatial regularity in radio-concentric networks, from very symmetric structures like amphitheaters (Figures~\ref{fig:europe} and \ref{fig:orange}, representing the European Parliament and the antique theater of Orange, respectively) or circular cities (Figure~\ref{fig:suncity}, which depicts a series of connected perfectly circular villages constituting SunCity), to more asymmetric urban shapes (Figure~\ref{fig:amsterdam}, showing the one-side radio-concentric shape of Amsterdam) or graphs with a more relaxed or degraded geometry (Figures~\ref{fig:paris} and \ref{fig:sfax}, presenting Paris and Sfax, respectively).

\subsection{Geographical Models of Rectilinear or Radio-concentric Shapes}
Networks are often studied in geography because they depict the visible human mark of population life in their territories. Beyond monographs which describe particular places, there exist a few typologies of urban networks and schemes generally explored and measured via topological structure or functional dimensions of the cities \cite{fol10,gen00}. Concerning the types of urban networks and graphs, Blanchard and Volchenkov \cite{blan09} presented a simple-faced classification of different types of route schemes, including rectilinear networks (e.g. Manhattan), “organic” towns (e.g. city of Bielefeld, North Rhine-Westphalia in Germany), shapes of “corals” (e.g. Amsterdam or Venice). In his book, Marshall elaborates different taxonomies of street patterns \cite{mar05}. However, it is also possible to design very theoretical networks in order to study their properties, other things being equal \cite{tho02}. Nevertheless, there is no consensual classification of the urban networks. Looking at the literature, it is interesting to notice that the publications about network design are actually shared by different disciplines involved in the field: (spatial) econometrics of transportation \cite{hend04}, mathematical optimization \cite{butt09}, information and communication technologies \cite{abb11} or social networks \cite{mac04}. These disciplines can be advantageously complemented by the domain of graph theory \cite{berg58,math07,barth11}.

On the one hand, the main contribution of the Manhattan networks in the domain of spatial modeling and measuring is the rectilinear distance calculation, which is related to the mathematical $L_1$-norm \cite{hak64}, compared to the Euclidian distance based on the $L_2$-norm \cite{joss13}. On the other hand, the radio-concentric framework generated several outstanding models. In 1924, E. Burgess proposed the concentric zone model to explain urban social structures \cite{Park1925}, followed by Hoyt in 1939, who defined the sector model based on the urban development along centrifugal networks \cite{Hoyt1939}. In parallel, W. Christaller \cite{chr33} and A. L\"osch \cite{los40} designed the central places theory to understand how cities are organized in territory, according to the distribution of goods and services to the population. These models deal with access to facilities and are based on network design and costs. In 1974, Perreur and Thisse defined the circum-radial distance based on such structures \cite{per74}. Tobler’s law \cite{Tobler1970} and the Newton gravity model also both indicate an inverse relation between the strength of a force $F_{ij}$ and the distance $d_{ij}$ separating two points $i$ and $j$ in geographical space, giving to the center(s) a particular status in networks. All these spatial theories participate in defining and emphasizing radio-concentric models and shapes; however they neither refute nor contradict rectilinear webs that can also own centers in different ways.

\section{Straightness in Theoretical Rectilinear and Radio-concentric Networks for Center-to-Periphery Routes}
In this section, we focus on the straightness of center-to-periphery routes, for both rectilinear and radio-concentric theoretical networks. The results are purely analytic, i.e. no simulation is involved. The networks are theoretical and perfectly regular, with pure geometric shapes. The proposed methods are specifically designed for these types of networks.

We call \textit{radius} an edge starting at the center of a radio-concentric network. An edge connecting two radii is called a \textit{side}. Unlike the rectilinear network, the radio-concentric network is controlled by a parameter $\theta$: the angle formed by two consecutive radii. The constraint on $\theta$ is that there must exist an integer $k$ such that $k = 2\pi/\theta$. We suppose that $k>2$, because with $k=1$ the sides are not defined, and with $k=2$, both radii are mingled. We call \textit{angular sector} the part of the unit circle between two consecutive radii. A \textit{half-sector} is the part of the unit circle between a radius and a neighboring bisector (cf. Figure~\ref{fig:radio-angle}).

\subsection{Definition of the Straightness}
\label{sec:StrDef}
For a pair of nodes, the \textit{Straightness} is the ratio of the spatial distance $d_S$ \textit{as the crow flies}, to the geodesic distance $d_G$ obtained by following the shortest path on the network:
\begin{equation}
	S = \frac{d_S}{d_G}
	\label{eq:Rectitude}
\end{equation}

This measures ranges from $0$ to $1$, a high value indicating that the graph-based shortest path is nearly straight, and contains few detours.

Coming from graph theory, this property is interesting in network assessment, because it measures a part (in a certain meaning) of the "accessibility" capacity of a network. It is a kind of relative efficiency to reach a point in a network. In our case, edges have neither impedance, nor direction. We do not consider any possible traffic jams in the flow propagation. In our assumption, speed is the same all over the graph and so time is proportional to distance.

\begin{figure}[!h]
	\centering
	\includegraphics[width=0.6\columnwidth]{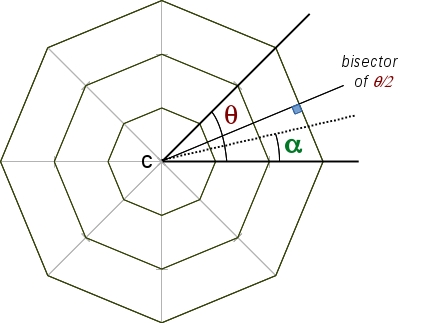}
	\caption{A perfect radio-concentric network, with $\theta$ separating two radial sections, and $\alpha$ in $[0,{\theta/2}]$..}
	\label{fig:radio-angle}
\end{figure}

Let the center of the network be the origin of a Cartesian coordinate system. In the rest of the document, we characterize a center-to-periphery move by an angle $\alpha$, formed by the $x$ axis and the segment going from the network center to the targeted peripheral node. The angle vertex is the network center, as represented in Figures~\ref{fig:manhattan-angle} and \ref{fig:radio-angle}. For the rectilinear network, we consequently note the straightness $S(\alpha)$ for the route of angle $\alpha$. For the rectilinear network, due to the presence of the parameter $\theta$ (the angle between two consecutive radii), we denote the straightness by $S_\theta(\alpha)$.

\subsection{Simplifying Properties of the Considered Networks}

\vspace{2mm}
\subsubsection{Rotation}
\begin{figure}[!h]
	\centering
	\includegraphics[width=0.5\columnwidth]{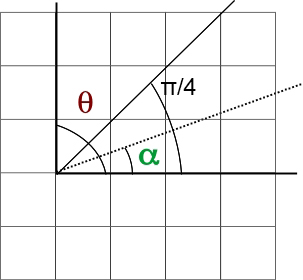}
	\caption{A perfect rectilinear network, with $\theta = \pi /2$ and $\alpha$ in $[0,{\pi/4}]$.}
	\label{fig:manhattan-angle}
\end{figure}

Both networks have certain \textit{rotation}-related properties, which allow some simplifications. As mentioned before, in a radio-concentric network, two consecutive radial sections of the network are separated by an angle $\theta$. In a rectilinear network (see Figure~\ref{fig:manhattan-angle}), a cell is a square. The angle between two consecutive edges originating from the network center is therefore $\theta = \pi /2$. So, we can distinguish $k=4$ angular sectors, corresponding to the quadrants of our Cartesian coordinate system. 

Both types of networks can be broken down to $k$ angular sectors, which are all similar modulo a rotation centered at the network center. So, \textit{without loss of generality}, we can restrict our analysis to the first angular sector, i.e. to the interval $\alpha \in [0;\theta]$.

\subsubsection{Homothety}
\begin{figure}[!h]
	\centering
	\includegraphics[width=0.8\columnwidth]{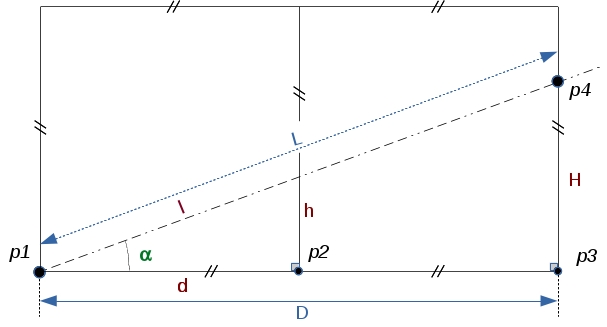}
	\caption{Geometry of a rectilinear network.}
	\label{fig:manhattan-homo}
\end{figure}

\begin{figure}[!h]
	\centering
	\includegraphics[width=0.8\columnwidth]{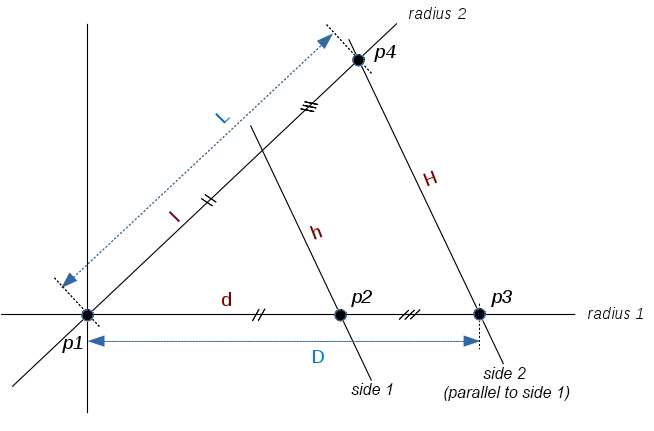}
	\caption{Geometry of a radio-concentric network.}
	\label{fig:radio-homo}
\end{figure}

An additional simplification comes from the \textit{homothetic} nature of both studied networks. Indeed, in these networks, a center-to-periphery route can go either through the edges originating from the network, or through those intersecting with these edges, as represented in Figures~\ref{fig:manhattan-homo} and \ref{fig:radio-homo}. Let us consider a move from $p_1$ to $p_4$. In both cases, $h$ is parallel to $H$, so we can deduce that:
\begin{equation*}
	\frac{l}{L} = \frac{d}{D} = \frac{h}{H}
\end{equation*}
Then we obtain:
\begin{equation*}
	l \cdot D = L \cdot d 
    \quad ; \quad 
    d \cdot H = D \cdot h 
    \quad ; \quad 
    l \cdot H = L \cdot h
\end{equation*}
We can set:
\begin{equation*}
	l \cdot D + l \cdot H = L \cdot d + L \cdot h
\end{equation*}
\noindent That is to say $S$ is the same whether the destination point is located on the closest or the farthest edge to the network center, for both types of networks:
\begin{equation}
	S = \frac{l}{d+h}=\frac{L}{D+H}
	\label{eq:Rectitude2}
\end{equation}

Consequently, \textit{without any loss of generality}, we can therefore restrict our analysis to the first mesh of the network, i.e. to the first square for the rectilinear network and to the first triangle for the radio-concentric network.

\subsubsection{Symmetry}
\begin{figure}[!h]
	\centering
	\includegraphics[width=1\columnwidth]{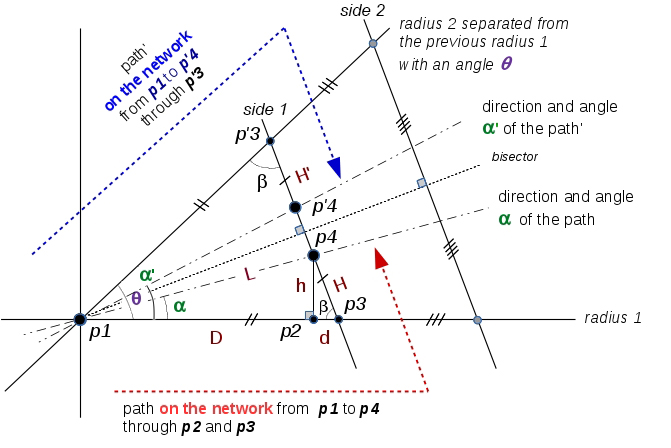}
	\caption{Basic paths followed on a radio-concentric geometric network.}
	\label{fig:radio-path}
\end{figure}

The last simplification comes from a symmetry property present in both networks, as illustrated by Figure~\ref{fig:radio-path} for radio-concentric networks. Observe that when $\alpha$ is smaller than $\theta / 2$, the shortest paths go through the first radius ($p_1p_3$), whereas as soon as $\alpha$ exceeds this threshold, they go through the second one ($p_1p'_3$). Also note that the line corresponding to this angle $\theta/2$ is the bisector of the angle formed by both radii.

Let us now consider the shortest path to $p_4$, which corresponds to an angle $\alpha < \theta / 2$. We wish to calculate the shortest path to some other point $p'_4$, corresponding to an angle $\alpha' > \theta / 2$, and such that the travelled distance is the same as for $p_4$. We know that $[p_1;p_3]$ and $[p_1;p_3']$ have the same length, so the distance to travel on the side is the same for both routes, i.e. $H = H'$. Moreover, the angles formed by each radius and the side are equal by construction: it is $\beta$. Let us now consider the triangles $(p_1,p_3,p_4)$ and $(p_1,p'_3,p'_4)$. We have two pairs of equal consecutive sides and the angles they form are equal; they are both $\beta$. So, both triangles are congruent, and we have: $\widehat{p_3,p_1,p_4} = \widehat{p'_3,p_1,p'_4}$. By definition, $\widehat{p_3,p_1,p_4} = \alpha$ and $\widehat{p_3',p_1,p_4} = \theta - \alpha'$, so we get $\alpha = \theta - \alpha'$, and finally $\alpha' = \theta - \alpha$. For the same reason (congruence), $L = L'$, meaning that the distances \textit{as the crow flies} are identical for $p_4$ and $p_4'$.

Since both distances (on and off the network) are the same for $p_4$ and $p'_4$, their straightness are also equal. In other words: $S_\theta(\alpha) = S_\theta(\theta - \alpha)$. We can conclude that, without any loss of generality, we can focus our study on the first half of the first angular sector, i.e. the interval $\alpha \in [0;\theta/2]$. The same proof can be applied to the rectilinear network, which displays the same type of symmetry. Consequently, the same simplification holds. 

\subsection{Analytic Expression of the Straightness}
In the previous section, we showed that, due to certain properties of rotation, homothety and symmetry, we can restrict our analysis of the straightness to only the interval $\alpha \in [0;\theta/2]$ for both networks. For the rectilinear network, note that $\theta = \pi/2$. So, in this specific case, we consider the interval $[0,\pi/4]$. Let us now give the expression of the straightness for each type of network. 

For the rectilinear network, we have from Figure~\ref{fig:manhattan-homo}:
\begin{equation*}
	\sin \alpha = \frac{h}{l}  \qquad ; \qquad \cos \alpha = \frac{d}{l}
\end{equation*}
Then, from~\eqref{eq:Rectitude2}, we obtain:
\begin{equation}
	S(\alpha) = \frac{l}{d+h} = \frac{1}{\cos \alpha +\sin \alpha}
	\label{eq:Rectilineaire}
\end{equation}

For the radio-concentric network, let us observe Figure~\ref{fig:radio-path}. Note that $p_2$ is the projection of $p_4$ onto the first radius. We obtain two rectangular triangles containing $p_2$: $(p_1;p_2;p_4)$ and $(p_4;p_2;p_3)$. This allows us to write the following equations:
\begin{equation*}
	\cos \alpha = \frac{D}{L} \qquad ; \qquad \cos \beta  = \frac{d}{H}
\end{equation*}
\begin{equation*}
	\sin \alpha = \frac{h}{L} \qquad ; \qquad \sin \beta = \frac{h}{H}
\end{equation*}
\begin{equation*}
	\tan \alpha = \frac{h}{D} \qquad ; \qquad \tan \beta = \frac{h}{d}
\end{equation*}
By removing $h$ we get
\begin{equation*}
	L \cdot \sin \alpha = D \cdot \tan \alpha = H \cdot \sin \beta = d \cdot \tan \beta,
\end{equation*}
and hence we obtain:
\begin{align*}
	D &= \frac{L \cdot \sin \alpha}{\tan \alpha}\\
    d &= \frac{L \cdot \sin \alpha}{\tan \beta}\\
    H &= \frac{L \cdot \sin \alpha}{\sin \beta}
\end{align*}
We substitute these values into~\eqref{eq:Rectitude}, then simplify and obtain:
\begin{align}
	S_{\theta}(\alpha) 
	&= \frac{L}{D+d+H} \\
	&= \frac{1}{\cos \alpha +\frac{\sin \alpha}{\tan \frac{\pi-\theta}{2}}+\frac{\sin \alpha}{\sin \frac{\pi-\theta}{2}}}
	\label{eq:Radioconcentrique}
\end{align}

Note that these formulas are valid only for the interval $\alpha \in [0;\theta/2]$. The other values can be deduced by symmetry and/or rotation, as explained earlier.

\subsection{Comparison of Networks}
\begin{figure}[!h]
	\centering
	\includegraphics[width=1\columnwidth]{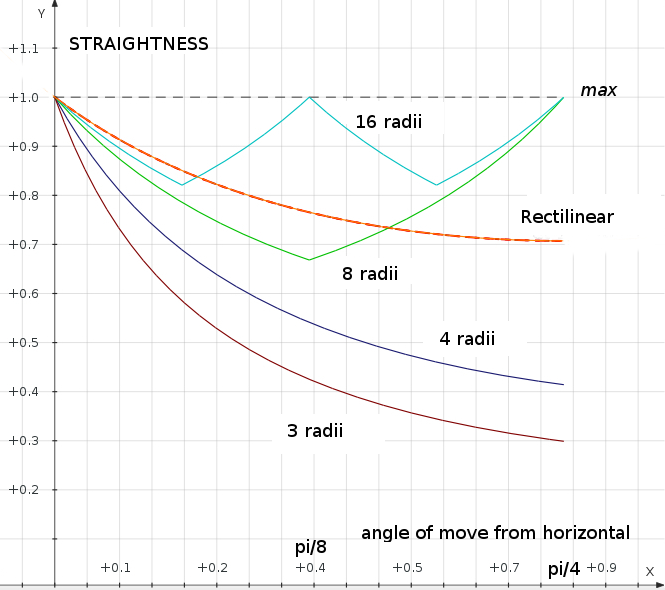}
	\caption{Straightness $S$ of different networks depending on the angle $\alpha$ of motion: $1$ rectilinear network and $4$ radio-concentric networks from $3$ to $16$ radii ; $\alpha \in \left[0;\frac{\pi}{4}\right]$.}
	\label{fig:comparison}
\end{figure}

With the analytic expression of the straightness for both types of networks, we can now compare their performance for center-to-periphery routes. Figure~\ref{fig:comparison} represents the straightness obtained for the different network types and parameter values. The $x$- and $y$-axis represent the angle $\alpha$ and the straightness $S$ calculated using the different previous formula, respectively. The optimal straightness value is represented by the black dotted horizontal line $f(\alpha)=1$. The closer the graphical representation of a network is to this line, the better the network is in terms of straightness.

For the radio-concentric network, we consider several values of the parameter $\theta$, corresponding to $k=3,4,8,16$, represented in purple, blue, green, and cyan, respectively. The rectilinear network is represented in red. The $x$-axis ranges only from $\alpha=0$ to $\pi/4$, which is enough, regarding the simplifications we previously described: we know all the plotted lines have a periodic behavior, which directly depends on $\theta$. 

For all lines, we have $S(0)=1$, which corresponds to a straightforward move on the first radius (or horizontal edge, for the rectilinear network). Then, the straightness decreases when $\alpha$ gets larger, since the destination point gets farther from this radius, and therefore from an optimal route. This corresponds to the lower route, represented in red in Figure~\ref{fig:radio-path}. The decrease stops when $\alpha$ reaches $\theta/2$ (i.e. the bisector): the straightness then starts increasing again, until it reaches $1$. This is due to the fact that the shortest path is now the upper route, represented in blue in Figure~\ref{fig:radio-path}. The maximal value is reached when $\alpha=\theta$, i.e. when the destination point lays on a radius, allowing for an optimal route. The same ripple pattern is then repeated again, and appears $k$ times.

As mentioned before, the periodicity directly depends on $\theta$: the smaller the angle, the larger the number of radii, which means the number of ripples increases while their size decreases. In other words, and unsurprisingly, the straightness of a radio-concentric network increases when its number of radii increases. More interesting is the fact that most of the time, $8$ (a very small number) radii are sufficient to make radio-concentric networks better than rectilinear ones.

\subsection{Boundary Condition}
Let us now consider a radio-concentric network with infinitely small $\theta$. From the definition of $k$, we know this would result in an infinitely large number of radii:
\begin{equation*}
	\lim_{\theta \to 0} k = \lim_{\theta \to 0} \frac{2\pi}{\theta} = + \infty
\end{equation*}

The radii of such a network would cover the whole surface of a disk. Since we focus our study on the first angular half-sector, $\alpha$ is itself bounded from above by $\theta/2$. So, if $\theta$ tends towards $0$, $\alpha$  does too. From~\eqref{eq:Radioconcentrique}, we consequently get:
\begin{equation}
	\lim_{\theta \to 0} S_{\theta}(\alpha) = \lim_{\theta \to 0} \frac{1}{\cos \alpha +\frac{\sin \alpha}{\tan \frac{\pi-\theta}{2}}+\frac{\sin \alpha}{\sin \frac{\pi}{2}-\frac{\theta}{2}}}
\end{equation}

We have: 
\begin{align*}
	\lim_{\theta \to 0} \cos \alpha &= \lim_{\alpha \to 0} \cos \alpha = 1 \\
    \lim_{\theta \to 0} \tan \frac{\pi-\theta}{2} &= + \infty \\
    \lim_{\theta \to 0} \sin \alpha &= \lim_{\alpha \to 0} \sin \alpha = 0 \\
	\lim_{\theta \to 0} \sin \frac{\pi-\theta}{2} &= 1
\end{align*}

We finally obtain the boundary for the straightness of a radio-concentric network with an infinite number of radii:
\begin{equation*}
	\lim_{\theta \to 0} S_{\theta}(\alpha) = 1
\end{equation*}

This result is obvious and consistent with the case where the complete route goes along a radius. Indeed, with an infinite number of radii, we can always find a radius to move directly from the center to the destination. Figure~\ref{fig:comparison} confirms this result: increasing the number of radii reduces the period and the amplitude of the \textit{ripples}, eventually leading to the optimal horizontal straight line of the equation $S=1$.

Between the two extreme cases of $3$ radii and an infinite number of radii, we should remind the reader that $8$ radii are enough for the radio-concentric network to become better than the rectilinear one in terms of straightness of center-to-periphery routes. Let us see now how it goes for multi-directional moves.

\section{Simulation of the Average Straightness for All Routes}
\begin{figure}[!h]
	\centering
	\includegraphics[width=0.5\columnwidth]{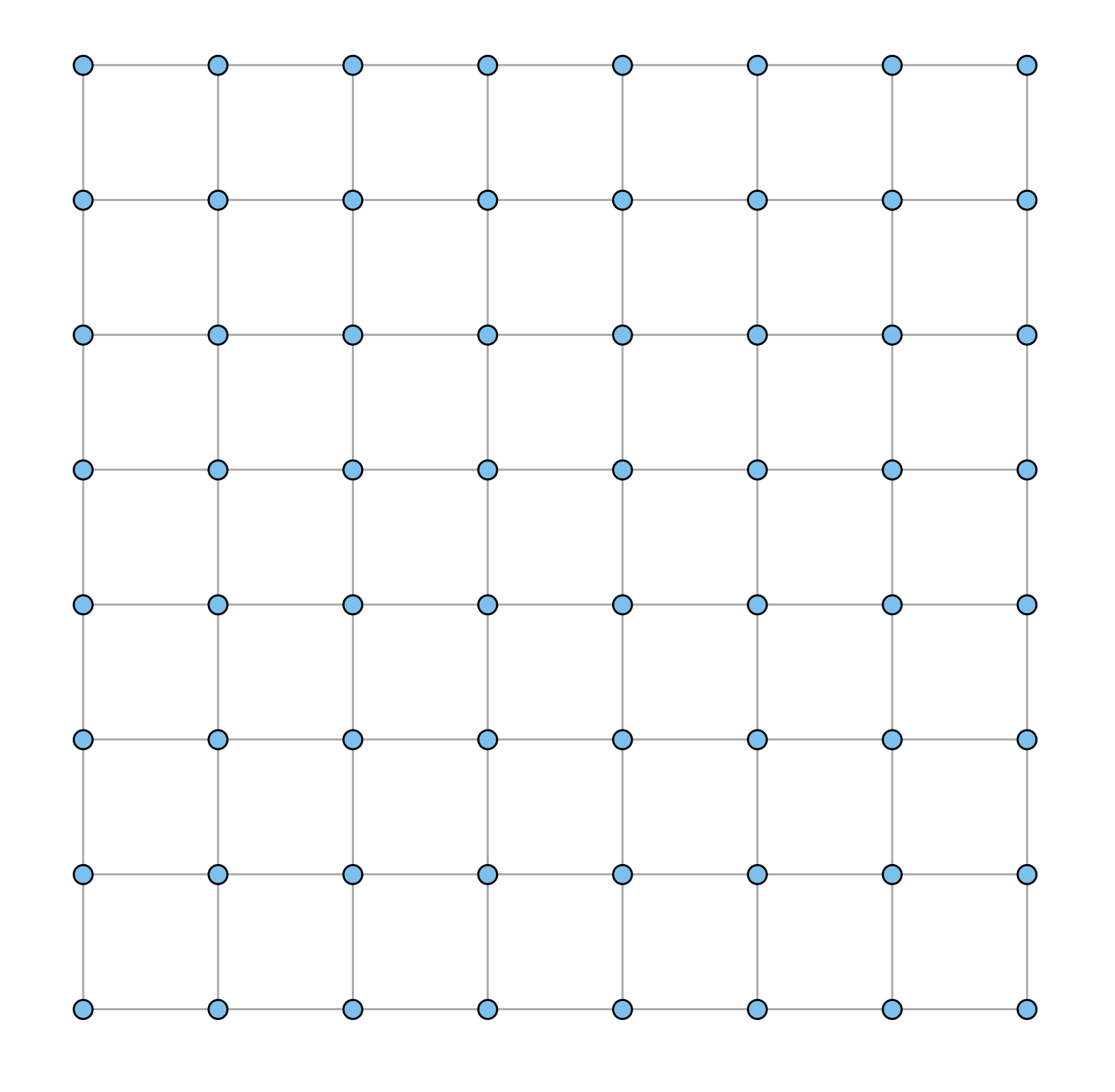}
	\caption{A perfect rectilinear graph.}
	\label{fig:recti-perfect}
\end{figure}

\begin{figure}[!h]
	\centering
	\includegraphics[width=0.6\columnwidth]{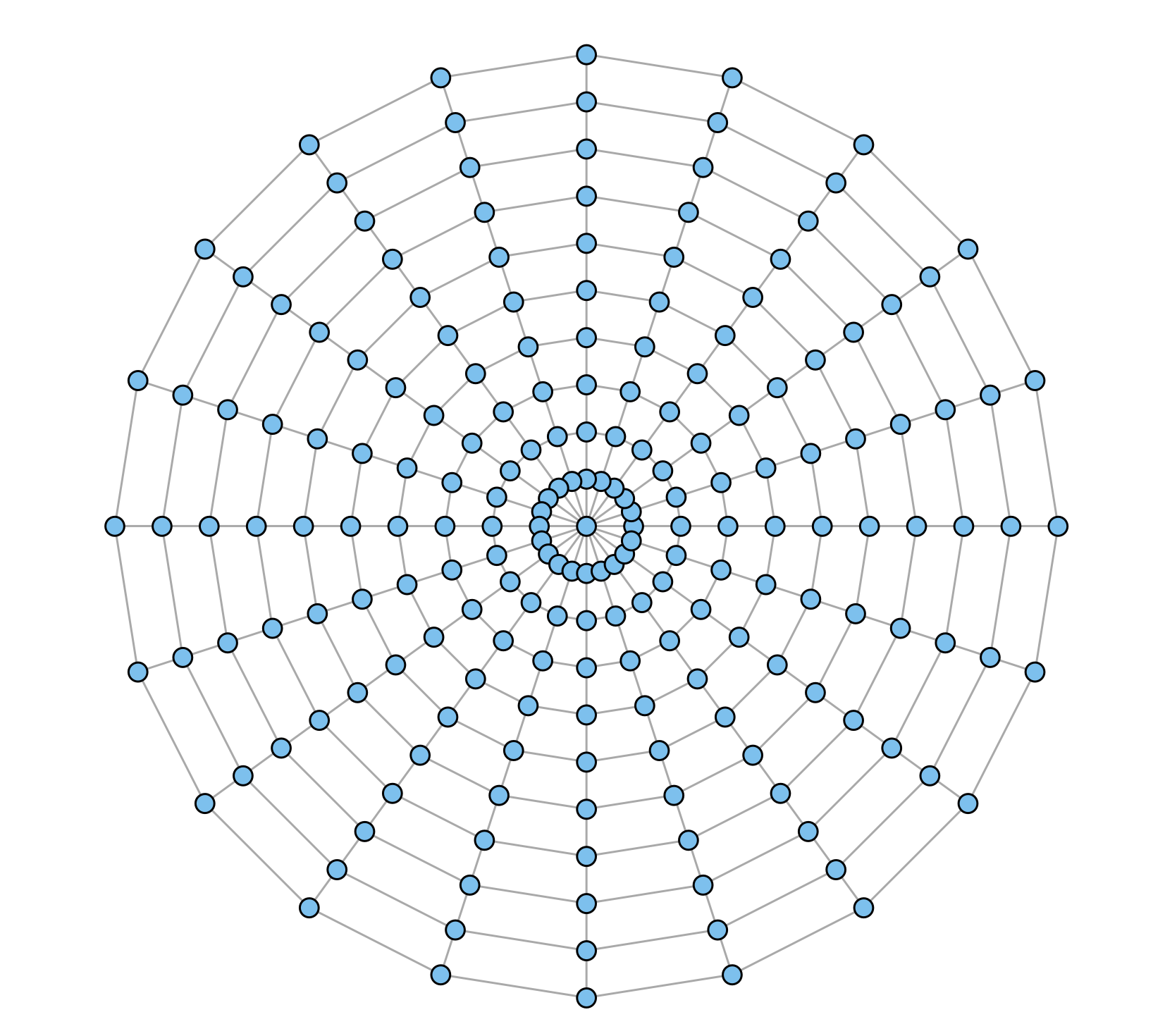}
	\caption{A perfect radio-concentric graph.}
	\label{fig:radio-perfect}
\end{figure}

Studying analytically the straightness for all possible moves (other than center-to-periphery) would be too time-consuming, so we switched to simulation. We used the statistical software R to simulate the motions on both rectilinear and radio-concentric networks. The shortest paths are processed using Dijkstra's algorithm \cite{dijk71}, for all pairs of nodes (see Figures~\ref{fig:recti-perfect} and \ref{fig:radio-perfect}).

\begin{figure}[!h]
	\centering
	\includegraphics[width=0.8\columnwidth]{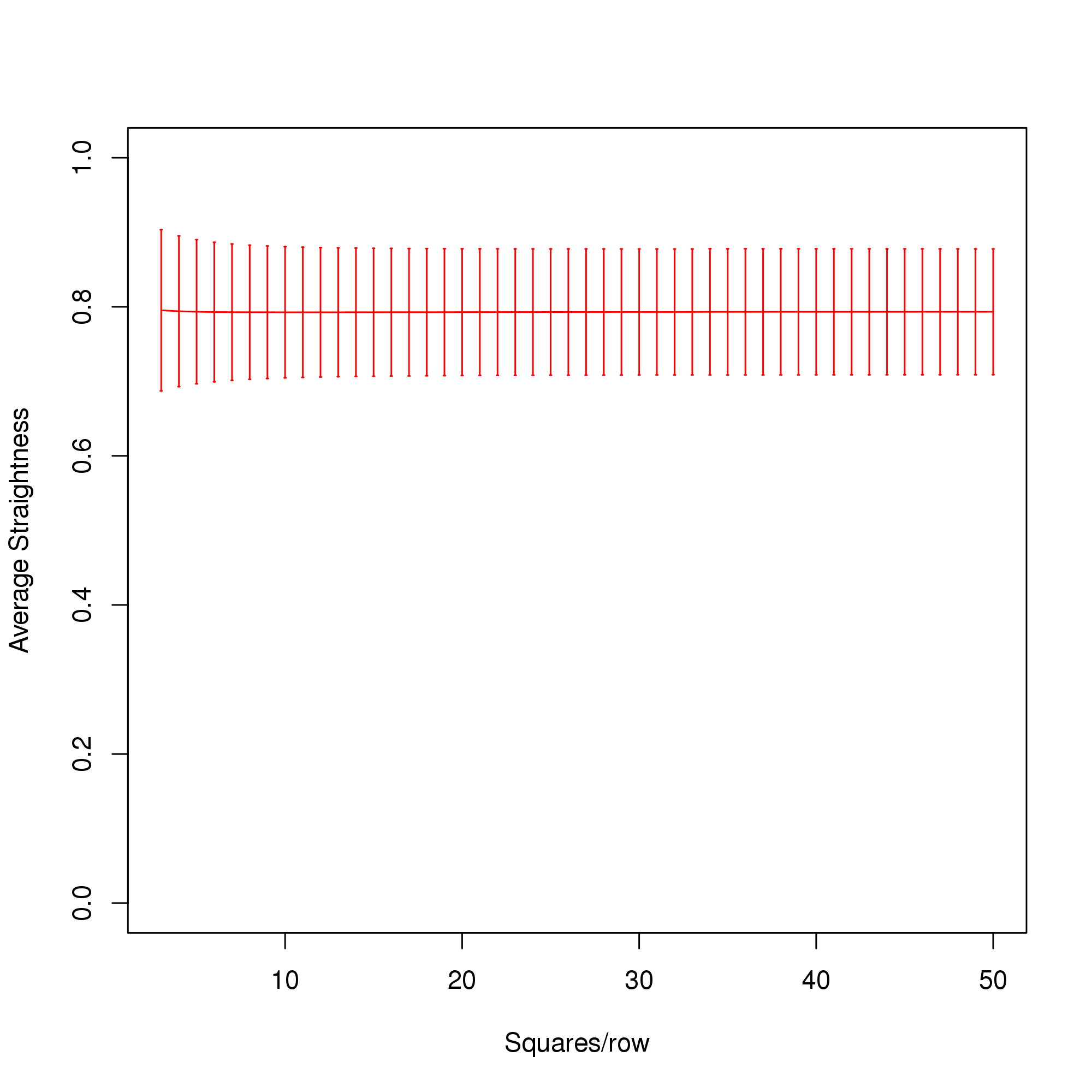}
	\caption{Average straightness $S$ (and standard deviation) for a rectilinear network, as a function of its size (expressed in number of squares by side).}
	\label{fig:average-straight-recti}
\end{figure}

These results are very interesting and confirm our theoretical findings, as well as our observations regarding the good straightness of radio-concentric networks, thrifty in number of radii, compared to rectilinear networks. Figure~\ref{fig:average-straight-recti} shows that, even when increasing the granularity of the grid forming a rectilinear network, the average straightness is constant, at a value lower than $0.8$. This is consistent with our remark regarding the homothety property of this network.

\begin{figure}[!h]
	\centering
	\includegraphics[width=0.8\columnwidth]{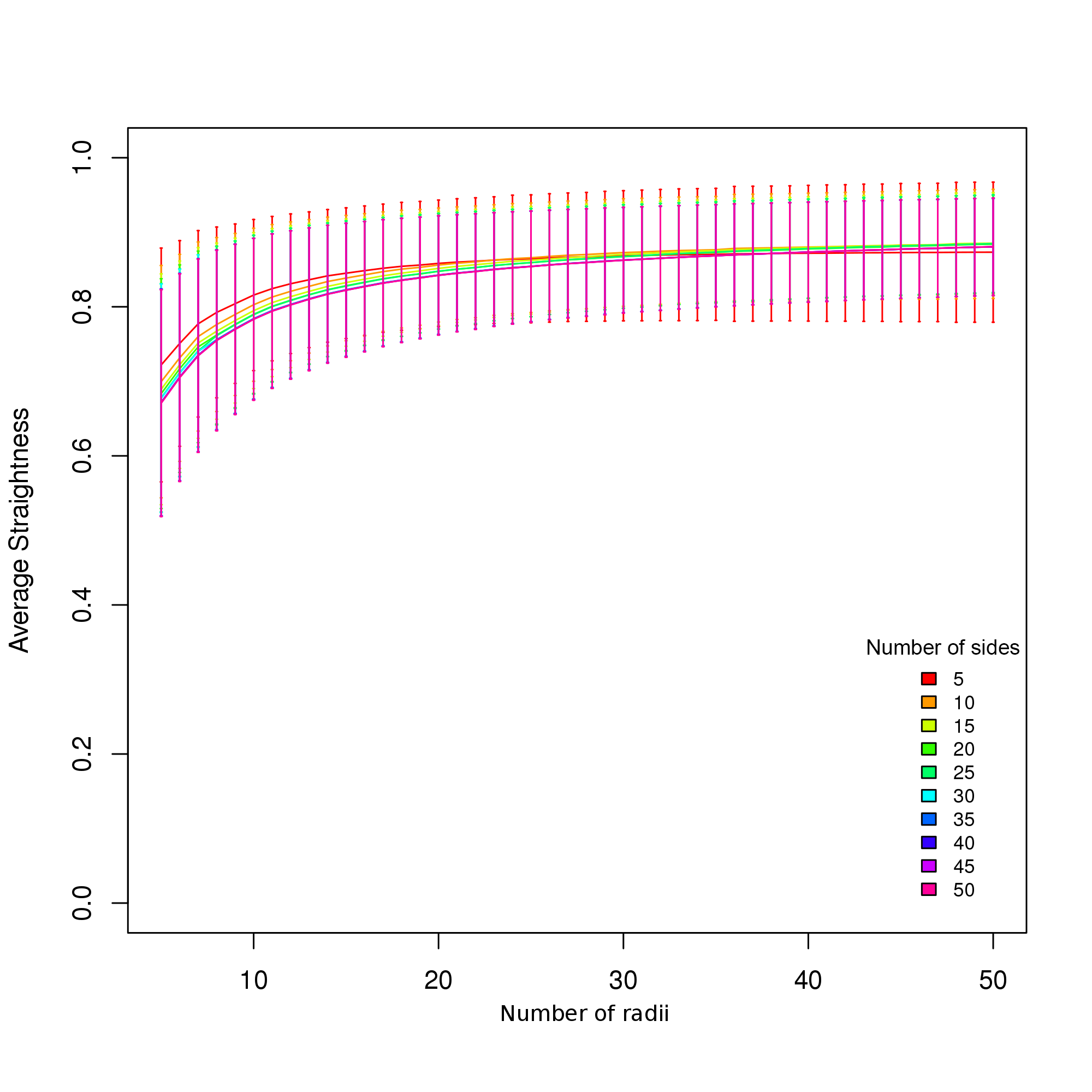}
	\caption{Average straightness $S$ (and standard deviation) for a radio-concentric network, as a function of its number of radii, and for different number of sides (see colors).}
	\label{fig:average-straight-radio}
\end{figure}

Regarding the radio-concentric network, Figure~\ref{fig:average-straight-radio} shows that the average straightness overtakes the rectilinear threshold ($0.8$) when the number of radii is about $8$ or $10$. It is interesting to notice that the number of sides does not affect the straightness much. This was expected for center-to-periphery routes, as for the rectilinear network. However, the routes considered here are more general, going from anywhere to anywhere, and moreover, the radio-concentric network is not a tessellation like the rectilinear one, so it is surprising to make this observation: further inquiry will be necessary to provide some explanations.

\section{Conclusion}
In this paper, we first presented a few urban networks based on rectilinear versus radio-concentric structures. Then, in a theoretical framework, we showed the superiority of the radio-concentric network compared to the rectilinear network, in terms of straightness. It was first demonstrated analytically in the particular case of center-periphery motions and then simulated on paths with multiple origins and destinations, using the statistical software R. To our knowledge, these results are new and original. They show that, straightness-wise, whatever the density of rectilinear networks, those cannot efficiently compete with radio-concentric networks, because their straightness is bounded by construction, at least within a theoretical framework.

However, this exploratory study does not take into account several factors that must be now studied to complete these first results. It will be interesting to find exactly over which number of radii a radio-concentric network has a better straightness than a rectilinear one. Knowing this threshold, we shall then be able to calculate the total length of both networks to fill an equivalent average straightness on similar surfaces to drain. What will be the most thrifty network in terms of length of "cables"? On another aspect, these calculations and simulations consider very theoretical networks. Geographers, architects and town planners may be interested in understanding the real straightness of urban, rectilinear versus radio-concentric networks, in real conditions (population mobility, congestion). This is also one of the further researches developed in the $Urbi\&Orbi$ project.

\phantomsection\addcontentsline{toc}{section}{Acknowledgments}
\section*{ACKNOWLEDGMENTS}
We would like to thank the CNRS (PEPS MoMIS) and the University of Avignon for supporting this research.

\balance

\phantomsection\addcontentsline{toc}{section}{References}
\bibliographystyle{plainnat}
\bibliography{biblio}

\end{document}